# Speckle from branching vasculature: dependence on number density

**Kevin J. Parker,**[a*] **Sedigheh S. Poul**[a]

[a]University of Rochester, Department of Electrical and Computer Engineering, 724 Computer Studies Building, Box 270231, Rochester, NY, 14727-0231, United States

[b]University of Rochester, Department of Mechanical Engineering, 235 Hopeman Engineering Building, Box 270132, Rochester, NY, 14727-0132, United States

**Abstract**. Recent theories examine the role of the fractal branching vasculature as a primary site of Born scattering from soft normal tissues. These derivations postulate that the first order statistics of speckle from tissue such as the liver, thyroid, and prostate will follow a Burr distribution with a power law parameter that can be related back to the underlying power law which governs the branching network. However, the issue of scatterer spacing, or the number of cylindrical vessels per sample volume of the interrogating pulse, has not been directly addressed. This is now examined with a 3D simulation that varies the number density and the governing power law parameter of an ensemble of different sized cylinders. The Burr distribution is found to be an appropriate model for the histogram of amplitudes from speckle regions, and the parameters track the underlying conditions. These results are also tested in a more general model of rat liver scans in normal vs. abnormal conditions, and the resulting Burr distributions are also found to be appropriate and sensitive to underlying scatterer distributions. These preliminary results suggest that the classical Burr distribution may be useful in the quantification of scattering of ultrasound from soft vascularized tissues.

*Kevin J. Parker**, E-mail: kevin.parker@rochester.edu

## 1 Introduction

The study of speckle statistics from ultrasound interrogation of soft tissues has a long and distinguished history with a number of treatments theories originally proposed in studies of radar and optical scattering from random materials and media. These treatments lead to classical models of the first order statistics of speckle, including Rayleigh distributions, homodyne-K, and marked regularity models[1-9]. The key assumptions about random acoustic scatterers within tissues have been commonly linked to cellular and connective tissues, furthermore linked to concepts such as average scatterer sizes and concentrations[10-22]. However, recently we proposed a different framework for analyzing backscatter from soft vascularized tissues such as the liver, prostate, brain, and thyroid. The key structures are considered to be the branching cylindrical network of fluid-filled channels that have a few percent difference in acoustic impedance from the reference media, which is the tissue parenchyma comprised of close packed cells. The branching fluid networks are self-similar, fractal networks and we hypothesize that the mathematical parameters of the fractal network substantially determine the first and second order statistics of backscatter[23-25] from these tissues.

To date, the effects of cylindrical vessel density (hyper-vascularized vs hypo-vascularized) and the effect of the size of the interrogated sample volume on the resulting first order statistics have not been examined. Simulations are conducted in k-Wave[26] to quantify the effects of these parameters. It appears that the first order statistics from a model of cylindrical branching vessels will vary smoothly within three different regimes: the sparse regime where there are few vessels per sample volume of the interrogating pulse, an intermediate range where the power law distribution of the vessels dominates, and then a high vessel density range where the first order statistics shift towards the classical Rayleigh (fully-developed) speckle. These results are compared with a few ultrasound exams from different tissues.

## 2   Theory

The first order statistics for the branching vasculature were recently derived[25]. Here we summarize the main points. First, we assume a broadband pulse propagating in the $x$ direction is given by separable functions:

$$P\left(y,z,t-\frac{x}{c}\right) = G_y\left(y,\sigma_y\right)G_z\left(z,\sigma_z\right)P_x\left(t-\frac{x}{c}\right). \tag{1}$$

For example, let $G_y\left(y,\sigma_y\right) = \exp\left[\left(-y^2/2\sigma_y^2\right)\right]$, i.e., Gaussian in the $y$ (and similarly in the $z$) direction, and where the pulse shape $P_x$ in the $x$ direction is given by:

$$P_x(x) = GH_2\left(\frac{x}{\sigma_x}\right)\exp\left[-\left(\frac{x}{\sigma_x}\right)^2\right] = e^{-x^2/\sigma_x^2}\left(\frac{4x^2}{\sigma_x^2} - 2\right), \tag{2}$$

where $GH_2$ is a second-order Hermite polynomial for the pulse shape with a spatial scale factor of $\sigma_x$[27,28], representing a broadband pulse. Its spatial Fourier transform is then:

$$^{3D}\mathfrak{I}\{P(x,y,z)\} = \left(4e^{-k_x^2\pi^2\sigma_x^2}k_x^2\pi^{5/2}\sigma_x^3\right)\left(e^{-2k_y^2\pi^2\sigma_y^2}\sqrt{2\pi}\sigma_y\right)\left(e^{-2k_z^2\pi^2\sigma_z^2}\sqrt{2\pi}\sigma_z\right), \tag{3}$$

where we use Bracewell's convention[29] for the form of the Fourier transform.

Using a 3D convolution model[2,30,31], we consider the *dominant* echoes from the pulse interacting with each generation of elements in a branching, fractal, self-similar set of vessels shown in **Fig. 1**, and whose number density follows a power law behavior $N(a) = N_0/a^b$. From these echoes, the histogram of envelopes is determined, by summing up over all the fractal branches.

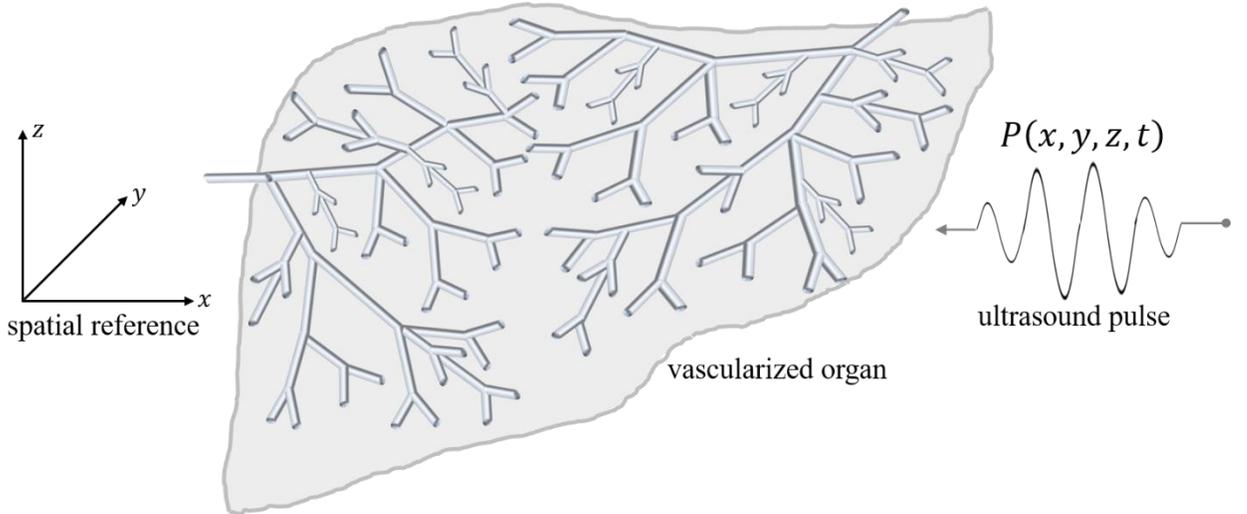

**Fig. 1** Model of 3D convolution of a pulse with the fractal branching cylindrical fluid-filled channels in a soft tissue.

The isotropic spatial and angular distribution of each generation of fractal branching structures is based on a scalable element. Specifically, consider a long fluid-filled cylinder of radius $a$:

$$f(r) = \begin{cases} \kappa_0 & r \leq a \\ 0 & r > a \end{cases}$$

$$F(\rho) = \frac{\kappa_0 \cdot a \cdot J_1[2\pi a \cdot \rho]}{\rho}, \quad (4)$$

where $\kappa_0$ is the fractional variation in compressibility, assumed to be $\ll 1$ consistent with the Born formulation, $F(\rho)$ represents the Hankel transform, which is the 2D Fourier transform of a radially symmetric function, $J_1[\cdot]$ is a Bessel function of order 1, and $\rho$ is the spatial frequency. The fractional variation in compressibility, $\kappa_0$, between blood vessels and liver parenchyma has been estimated to be approximately 0.03, or a 3% difference based on published data[24].

The convolution of the pulse with a cylinder of radius $a$ is dominated by the case where the cylinder is perpendicular to the direction of the forward propagating pulse, the $x-$axis in our case. Thus, assuming an optimal alignment, the 3D convolution result is given by the product of the transforms:

$$^{3D}\Im\{\text{echo}(x,y,z)\} = \Im^{3D}\{p(x,y,z)\} \bullet (k_x)^2 \Im^{3D}\{cylinder(x,y,z)\} =$$
$$\left(\left(4\mathbf{e}^{-k_x^2\pi^2\sigma_x^2}k_x^2\pi^{5/2}\sigma_x^3\right)\left(\mathbf{e}^{-2k_y^2\pi^2\sigma_y^2}\sqrt{2\pi}\sigma_y\right)\left(\mathbf{e}^{-2k_z^2\pi^2\sigma_z^2}\sqrt{2\pi}\sigma_z\right)\right) \bullet \quad (5)$$
$$(k_x)^2 \kappa_0 \left( a \left[\frac{1}{\sqrt{k_x^2+k_y^2}}\right] \bullet \left(J_1\left[2a\pi\sqrt{k_x^2+k_y^2}\right]\right)\right)\delta[k_z],$$

where the $(k_x)^2$ term pre-multiplying the cylinder transform stems from the Laplacian spatial derivative in the Born scattering formulation[32,33] and in the 3D convolution model[2,34].

By Parseval's theorem, the integral of the square of the transform equals the integral of the square of the echo, and after integration over the delta function in $k_z$:

$$\iiint \{\text{echo}(x, y, z)\}^2 \, dx dy dz = \sigma_z^2 \kappa_0^2 \int_{kx=-\infty}^{\infty} \int_{ky=-\infty}^{\infty} \left( 8\mathrm{e}^{-\pi^2 \left( k_x^2 \sigma_x^2 + 2 k_y^2 \sigma_y^2 \right)} k_x^2 \pi^{7/2} \sigma_x^3 \sigma_y \right)^2 \cdot$$
$$\left( k_x^2 \right)^2 \left( a \left[ \frac{1}{\sqrt{k_x^2 + k_y^2}} \right] \cdot \left( J_1 \left[ 2 a \pi \sqrt{k_x^2 + k_y^2} \right] \right) \right)^2 dk_x dk_y. \tag{6}$$

The square root of this gives the root mean square (RMS) amplitude of the echo, as a function of the radius $a$. We will associate the RMS amplitude from each echo with a proportionally higher maximum value of the envelope, as a function of cylinder radius $a$, denoted as $A[a]$.

From this framework and numerical evaluation of eqn (6) we found[25] an approximation which will be useful for deriving a closed form solution of the echo amplitude, $A[a] = A_0 \sqrt{a - a_{\min}}$, justified by the nearly linear increase in the energy term above some minimum threshold, and the asymptotic modulus of $J_1(ak)$ which is proportional to $\sqrt{2/(\pi ak)}$ [35] as $ak$ becomes large. Of course the exact shape is dependent on the particular pulse shape's spectrum and the beampattern.

So as a general approximation, we apply the relation $A[a] = A_0 \sqrt{a - a_{\min}}$ for $a > a_{\min}$. The parameter $a_{\min}$ depends on a number of factors, including the dynamic range selected (for example, 45 dB) and the Rayleigh scattering (long wavelength, small $a$) behavior of the cylinder interacting with the particular pulse transmit signal, along with the noise floor and quantization floor of the receiver.

Now, applying the general theory of transformed distributions[36], we have within the ensemble the number density of vessels at different radii given by $N[a] = N_0/a^b$, and this will be transformed into the distribution of amplitudes, $A(a)$. The general rule is:

$$N[A] = \frac{1}{dA/da} N[a]. \tag{7}$$

In our case, the derivative $dA/da = [(1/2)A_0]/\sqrt{a - a_{min}}$, and the inverse function is $a[A] = (A/A_0)^2 + a_{min}$. Thus, substituting these into eqn (7) the distribution $N[A]$ is:

$$N[A] = \frac{2N_0 A}{A_0^2 \left[ (A/A_0)^2 + a_{min} \right]^b}. \tag{8}$$

So, for example, if $b = 2$ and $A_0$ and $N_0$ are unity, then $N[A] = 2A/(A^2 + a_{min})^2$, and this is plotted in **Fig. 2** along with variations in parameters.

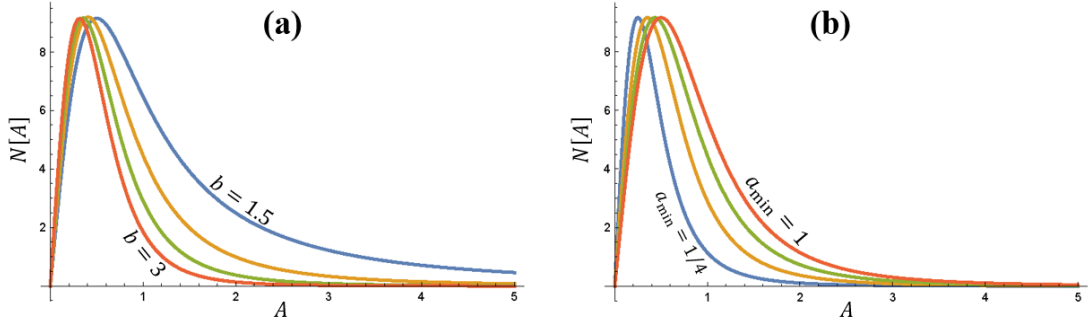

**Fig. 2** The proposed histogram function of envelope amplitudes $A$, having the form $A/(A^2 + a_{min})^b$. In (a) are normalized functions where $a_{min} = 1/2$ and the power law parameter $b$ is 3, 2.5, 2, and 1.5. In (b) are normalized functions where the power law parameter is fixed at 2.5 however $a_{min}$ is varied as 1/4, 1/2, 3/4, and 1. Vertical axis: counts (arbitrary units); horizontal axis: envelope amplitude (arbitrary units).

This provides a four-parameter fit $\{N_0, A_0, a_{min}, b\}$ to a histogram taken from a reasonably sized region of interest (ROI) within a vascularized tissue or organ, assuming an isotropic and spatially uniform distribution across the ROI.

However, of these four parameters, $N_0$, $A_0$, and $a_{min}$ are influenced by system parameters such as amplifier gain and the size of the ROI. To simplify the analysis, one can normalize by the integral of the distribution $\int N[A]dA = N_0 \big/ \big[(b-1)(a_{min})^{b-1}\big]$ to form a proper probability density function (PDF), which integrates to unity:

$$N_n[A] = \frac{2A(a_{min})^{b-1}(b-1)}{A_0^2\left[\left(\frac{A}{A_0}\right)^2 + a_{min}\right]^b}. \tag{9}$$

Furthermore, by substituting $\lambda = A_0\sqrt{a_{min}}$, we find this reduces to a two-parameter distribution:

$$N_n[A] = \frac{2A(b-1)}{\lambda^2\left[\left(\frac{A}{\lambda}\right)^2 + 1\right]^b}, \tag{10}$$

which is a Burr Type XII distribution[37,38] with $c = 2$. Thus, the normalized distribution offers a simplification to a two-parameter distribution with analytic expressions for probability density function, cumulative distribution function, and moments[38]. For example, the peak of the distribution occurs at $A = \lambda/\sqrt{2b-1}$ for $b > 1/2$.

Thus, we argue that the Burr distribution (10) is the expected histogram distribution of echo amplitudes from a fractal branching set of Born cylinders. In particular, the power law parameter $b$ is a major parameter of interest.

## 3   Methods

The numerical simulations are achieved using k-Wave to simulate the time-domain propagation of compressional wave in 3D. k-Wave is an open-source toolbox developed in MATLAB that solves the acoustic wave equations using the k-space pseudospectral method.[26]

In this study, the simulation domain is a 3D block of 15 mm (in depth, $x$) × 13 mm (in the lateral direction, $y$) × 3 mm (in the transverse direction, $z$). It is uniformly divided into small grid elements of approximately 69.4 $\mu$m in the $x$, $y$, and $z$ directions. The 3D orientation of the transducer for simulation is shown in **Fig. 3(a)**. A few cylindrical branches are also shown here for the clarification of the random branching orientation in the domain.

The medium consists of a uniform background and a set of cylindrical branches with different radii ranging from 1 to 6 grid elements mimicking the scattering structures as vessels. These branches are distributed randomly in the domain using the uniform random distribution function in MATLAB, and our algorithm assures that there is no overlap among any two branches generated. The cylindrical branches are placed along the transverse direction ($z$) perpendicular to the direction of propagation ($x$) since the orientation provides the dominant echoes accounted for in our theory.[25] The round cylindrical shapes in the model are not exact due to the discretization of radii to grid elements; flat surfaces may appear in some areas on the surface of cylindrical branches. Since flat walls are not realistic and result in strong reflective surfaces, the cylindrical radii are perturbed randomly to avoid the effect of artificially flat boundaries.

Each cylindrical branch with radius $a$ has a number density of $N(a)$ in the medium prescribed by the fractal branch power law relation as $N(a) = N_0/a^b$. In this study, three sets of different simulations are performed, each based on the value that is picked for the power law parameter $b$ from the physiological region of interest ($2 < b < 3$): $b = 2.2$, 2.5, and 2.8. For each set, $N_0$ is varied from 100 to 400 in increments of 50, guaranteeing that for each $N_0$ there is at least one cylinder with the largest radius using the power law formulae. In **Fig. 3(b)**, a 2D representation of cylindrical branch distributions for the case of $b = 2.8$ and 200 is shown.

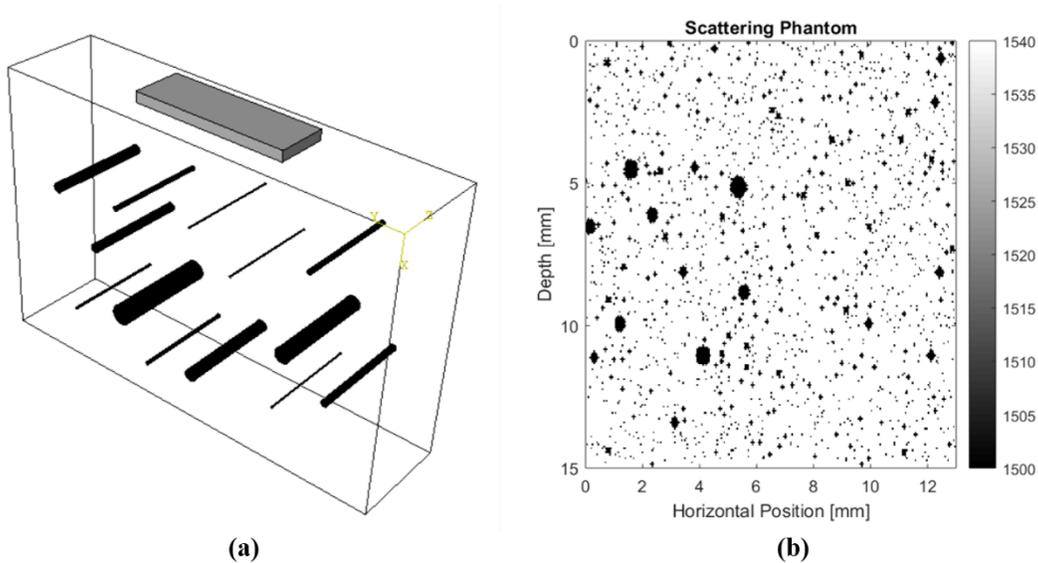

(a)            (b)

**Fig. 3** (a) 3D orientation of the transducer in the simulation domain. A few cylindrical branches are also shown here to clarify the random branches orientation in the domain. The length of the transducer shown here is relatively longer than that of simulation. (b) 2D view $(x - y)$ of the random distribution of cylindrical scatterers shown as black circular spots in the uniform white background, corresponding to the case of $b = 2.8$ and $N_0 = 200$.

Two cycles with a frequency of 4 MHz are applied as the toneburst excitation signal using a virtual linear array transducer defined in k-Wave that serves as both source and sensor to the transmit signal and the receive reflection signal, respectively. The transducer is focused at a depth of 10 mm from the top surface of the domain. The center frequency of 4 MHz is applied to mimic the frequency used in tissue scans. Other physical properties of the transducer are listed in **Table 1**. To avoid side lobe effects, the element width of the transducer satisfies the following condition:

$$\text{element width} \leq \frac{\lambda}{2}, \tag{11}$$

where $\lambda$ is the wavelength.

**Table 1** Physical properties of the transducer in the simulation.

| Transducer properties | Value |
|---|---|
| Number of elements | 32 |
| Focus | 10 mm |
| Element width | 0.0694 mm |
| Element height | 1.597 mm |
| Elevation focus | 10 mm |
| Kerf | 0 |

The properties of the medium defined here are speed of sound, density and absorption coefficients. As the medium is heterogeneous, these are defined as matrices in the size of the whole computational domain and a value is assigned to each single element. For elements in the background and cylindrical branch regions, the speed of sound is set to 1540 m/s and 1500m/s, respectively. The density is assumed to be uniform in the whole medium, taking the value of 1000 kg/m$^2$. The absorption coefficient is set to a small value.

Computational time step size is set using a Courant-Friedrichs-Lewy (CFL) number smaller than 0.3 to make the simulation stable $(\text{CFL} = c_0 \ dt/dx)$.[39] The B-mode image of the domain is reconstructed using sum and delay beamforming, frequency filtering, envelope detection, and log compression from the raw RF data.

Separately, experimental results were obtained from liver experiments. Rat experiments were reviewed and approved by the Institutional Animal Care and Use Committee of Pfizer, Inc. Groton Connecticut, where the ultrasound scan was acquired using a Vevo 2100 (VisualSonics, Toronto, CA) scanner and a 20 MHz center frequency transducer (data provided courtesy of Terry Swanson). Parameter estimation was performed using MATLAB (The Mathworks, Inc., Natick, MA, USA) nonlinear least squares minimization of error, for two-parameter fits of the Burr distribution to the data.

## 4 Results

A range of simulations was evaluated with the cylindrical scatterer density parameters varied between a power law $b$ of 2.2, 2.5, and 2.8 and $N_0$ of 100, 200, 300, and 400. Visually, the results shown in **Fig. 4** demonstrate the higher density of scatterers with increasing $N_0$. The envelope of the return echoes from these different simulations indicate progressive shifts as the number density of scatterers changes.

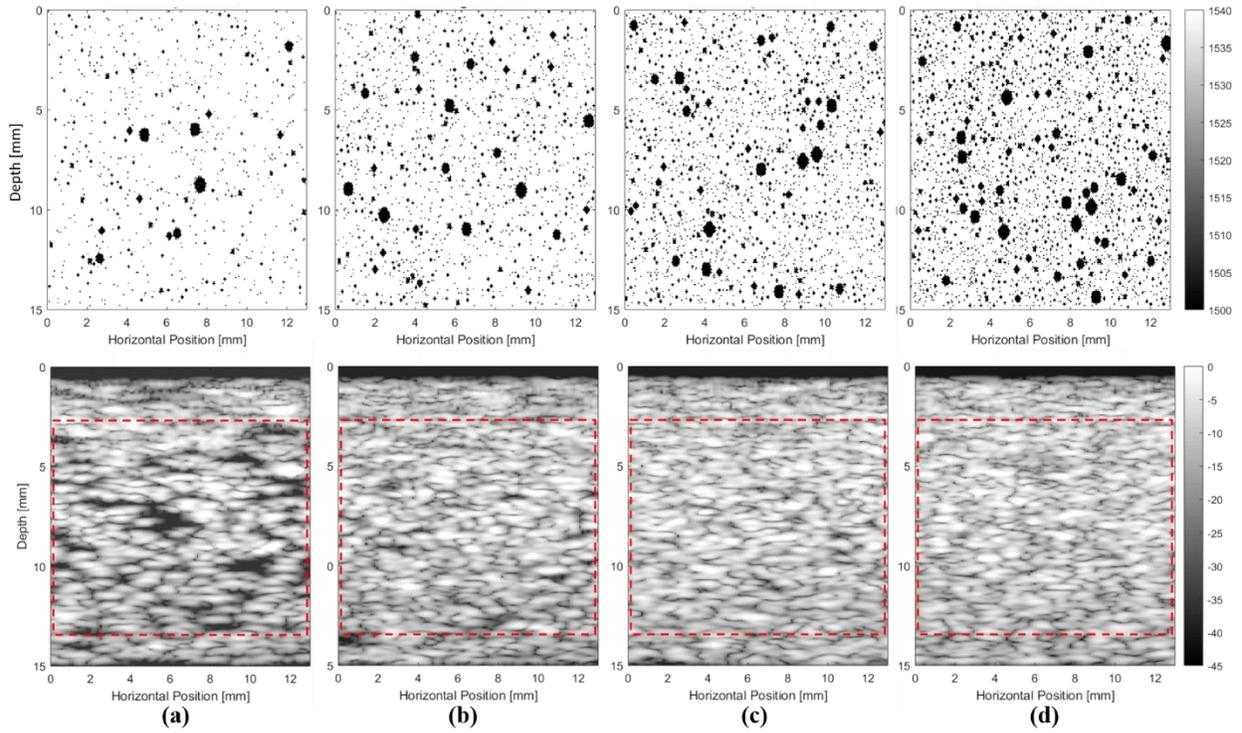

**Fig. 4** Comparison of the random scattering fields and corresponding B-mode images for different number densities based on the power law equation for $b = 2.5$. Column (a) $N_0 = 100$, column (b) $N_0 = 200$, column (c) $N_0 = 300$, and column (d) $N_0 = 400$. A region of interest (dashed lines) is shown for analysis.

In **Fig. 5** are single A-line envelopes taken from the center vertical line in the simulations of **Fig. 4**. At the lowest scatterer density, the envelope has many regions at or near zero. However, as the density increases, the behavior indicates more and higher local maxima with sharp speckle minima that indicate complex summations of nearby scattering sites.

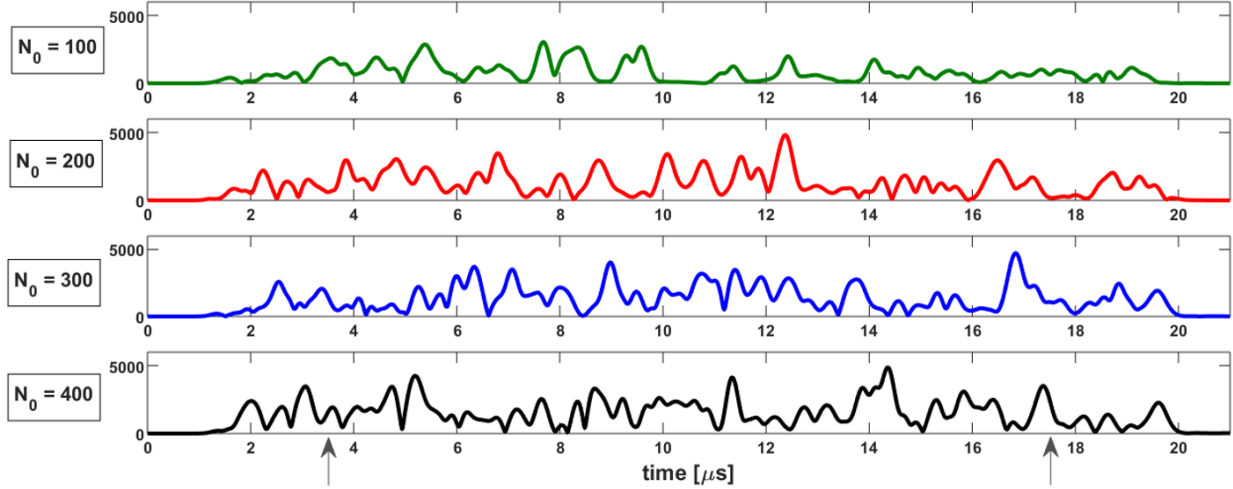

**Fig. 5** Envelope plots along the middle line for the power law parameter $b = 2.5$. The two arrows indicate the length of the ROI from which the histogram plots are obtained.

The histograms for three of these cases are shown in **Fig. 6.** The case of $N_0 = 100$ is not shown because the excess amount of signal at or near zero creates a poor curve fit to eqn (10); large anechoic regions are not accounted for in the theory.

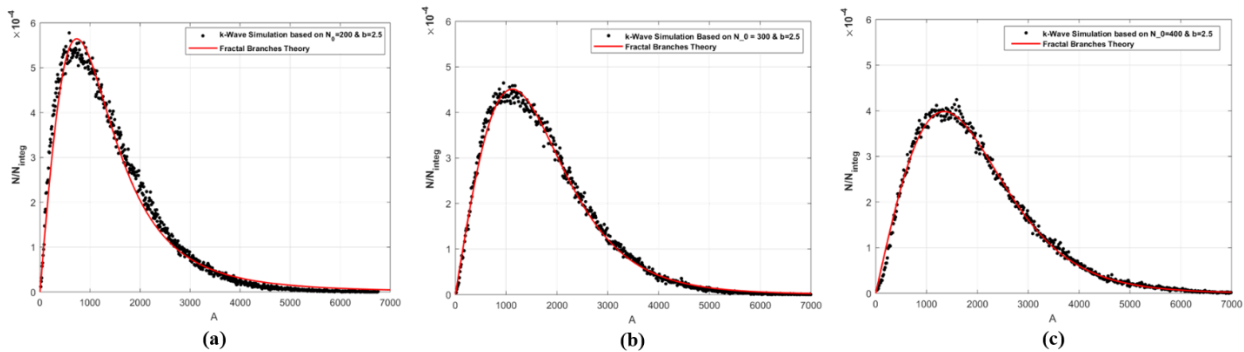

**Fig. 6** Fractal branches curve-fitting for: (a) $N_0 = 200$ and $b = 2.5$. Curve-fitting parameters: $\hat{b} = 2.341$, $\lambda = 1412$. Goodness of fit: $R^2 = 0.9869$, RMSE = 0.2088. (b) $N_0 = 300$ and $b = 2.5$. Curve-fitting parameters: $\hat{b} = 4.048$, $\lambda = 2980$. Goodness of fit: $R^2 = 0.9954$, RMSE = 0.1059. (c) $N_0 = 400$ and $b = 2.5$. Curve-fitting parameters: $\hat{b} = 5.593$, $\lambda = 4280$. Goodness of fit: $R^2 = 0.996$, RMSE = 0.08801.

A summary of all results over the parameter space is given in **Fig. 7**, also indicating the range of results found over 10 independent simulations of identical parameters. The clear trend is for an increasing curve fit of the power law estimate $\hat{b}$ with the generating power law $b$, but also with the number density of scatterers. This may be explained by **Table 2** which reports the average number of scattering cylinders that would lie within the volume of the interrogated pulse. The sample volume produced by the 4 MHz pulse was found to be roughly elliptical with a -10 dB area of approximately 0.26 mm$^2$, averaged over the ROI. When the average is less than 1 there are significant anechoic regions, however as the scatterer density per sample volume increases towards 4 or higher, significant complex summations can result, and higher estimates of the power law $\hat{b}$ are obtained from the histogram of the envelopes.

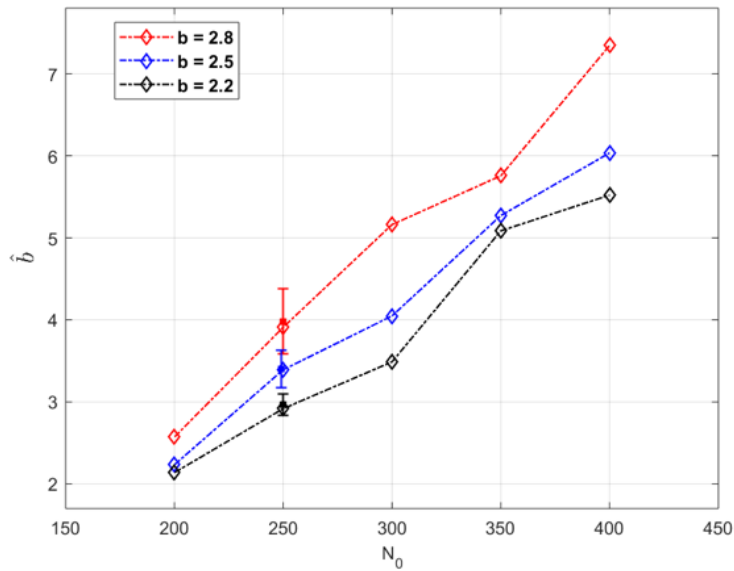

**Fig. 7** Summary of simulation results using the fractal branches theory for comparison of results for fitted $\hat{b}$. Error bars for fitting parameter are also shown when $N_0 = 250$ and $b = 2.2,\ 2.5,$ and 2.8, each resulting from 10 repetitions of the simulation.

**Table 2** Averaged number of scatterers within pulse area in k-Wave simulation.

| Variable | | $N_0 = 100$ | $N_0 = 200$ | $N_0 = 300$ | $N_0 = 400$ |
|---|---|---|---|---|---|
| $b = 2.2$ | Number of cylinders/pulse | 0.81 | 1.62 | 2.43 | 3.25 |
| $b = 2.5$ | Number of cylinders/pulse | 0.94 | 1.88 | 2.82 | 3.77 |
| $b = 2.8$ | Number of cylinders/pulse | 1.11 | 2.21 | 3.32 | 4.43 |

To test the generality of the results, we examine *in vivo* scans from rat liver experiments. A normal liver is shown in **Fig. 8** with histogram and fit to a Burr distribution ($\hat{b} = 2.8$), then in **Fig. 9** a fibrotic liver with low fat ($\hat{b} = 2.1$), and finally in **Fig. 10** a fibrotic liver with high fat accumulation in vesicles ($\hat{b} = 3.2$). These examples are not simple manipulations of cylindrical number density, and so represent more general cases than those shown in simulations. However, the fibrotic mesh can be considered to be an increase (compared to normals) of larger scatterers, hence consistent with a lower $b$ (consistent with a number density favoring the larger scatterers). The case with fibrosis and significant accumulation of microvesicular and macrovesicular fat would be consistent with a higher $b$ (favoring a high density of smaller scatterers). Thus, the results indicate some degree of generality of the framework of the Burr distribution, not restricted solely to cylindrical scatterers.

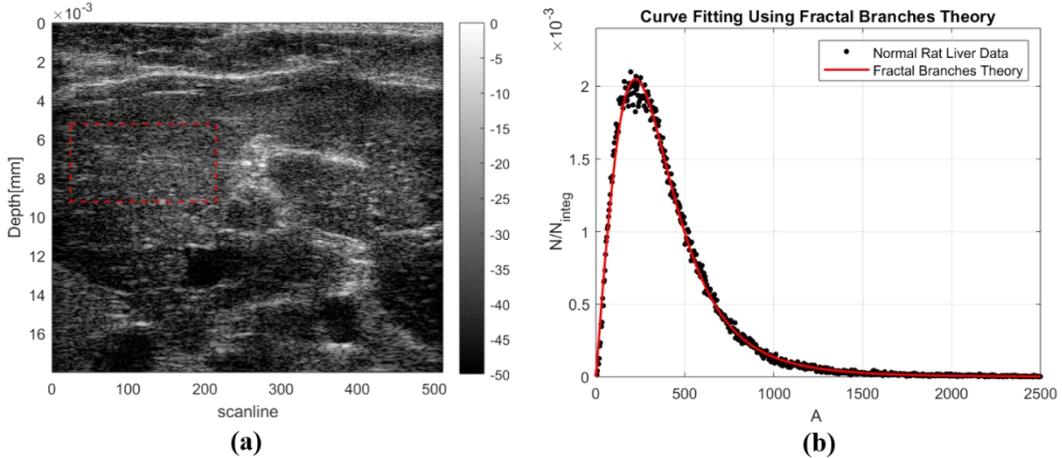

Fig. 8 (a) B-scan image of normal rat liver. A region of interest is selected (dashed lines) for analysis. (b) Fractal branching theory fitting to the histogram of normalized echo amplitude. Curve-fitting parameters: $\hat{b} = 2.832$, $\lambda = 407.1$. Goodness of fit: $R^2 = 0.996$, RMSE = 0.04665.

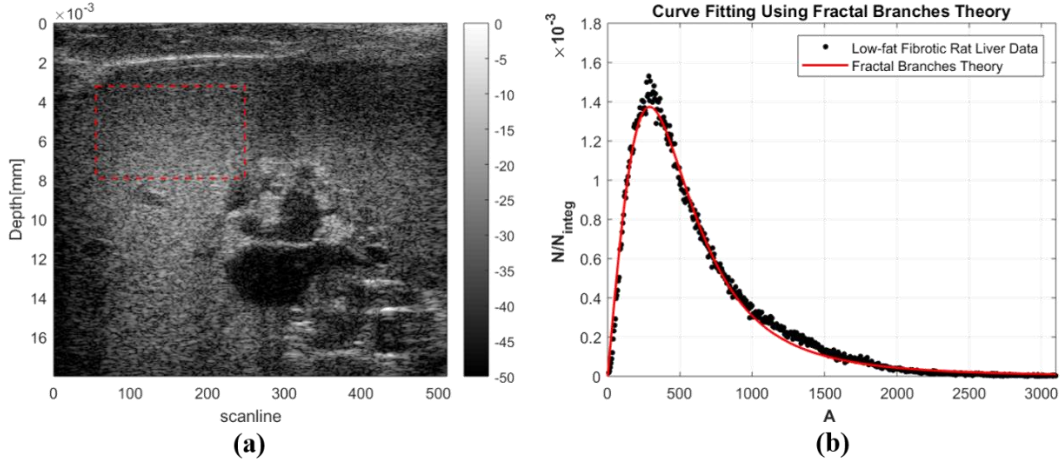

**Fig. 9** (a) B-scan image of low-fat fibrotic rat liver. A region of interest is selected (dashed lines) for analysis. (b) Fractal branching theory fitting to the histogram of normalized echo amplitude. Curve-fitting parameters: $\hat{b} = 2.184$, $\lambda = 532.5$. Goodness of fit: $R^2 = 0.9933$, RMSE = 0.03381.

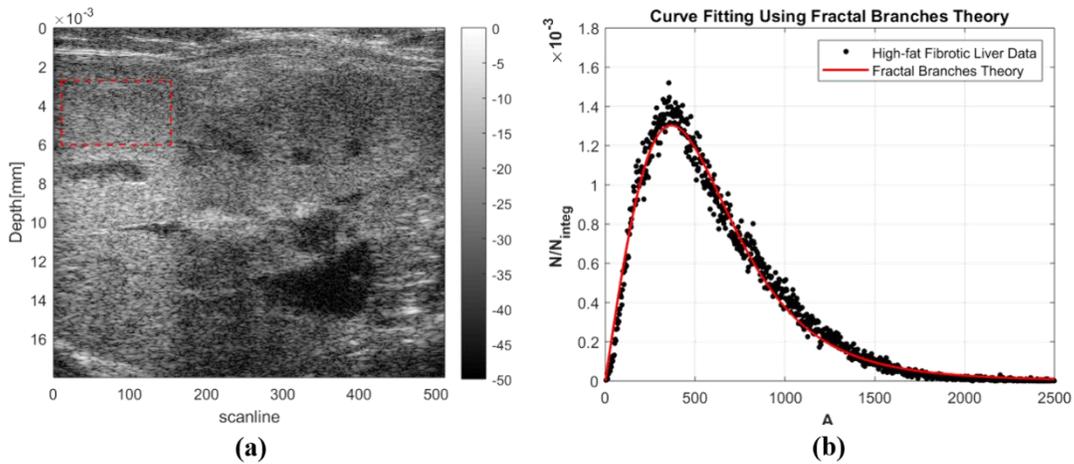

**Fig. 10** (a) B-scan image of high-fat fibrotic rat liver. A region of interest is selected (dashed lines) for analysis. (b) Fractal branching theory fitting to the histogram of normalized echo amplitude. Curve-fitting parameters: $\hat{b} = 3.219$, $\lambda = 871.8$. Goodness of fit: $R^2 = 0.9878$; adjusted RMSE = 0.04887.

## 5 Discussion and Conclusion

The derivation leading to the Burr distribution for speckle[25] has a number of key assumptions; whereby the accounting of echo amplitudes presumes that each cylindrical vessel at normal incidence to the interrogating pulse contributes according to its own scattering transfer function.

There is no accounting for multiple vessels within the sample volume formed by the interrogating pulse, and when these conditions are present the estimated $\hat{b}$ from the Burr distribution should approximate the underlying power law that governs the fractal branching of the vascular tree. From the numerical simulations we see this is valid within a zone of scatterer density of between one and two average cylindrical scatterers per sample volume of the interrogating pulse. Below that zone there are few scatterers and much of the envelopes from a scan line are at or near zero from the asymptotic decay of the echo at long distances from the rare scatterers. This skews the distribution towards lower values of echo amplitudes.

Conversely, when the number density of vessels is so high that there are many within a pulse's volume, then complex summation becomes important. In the classical theory where the scatterers were identical and at a number density higher than approximately seven[40], then we would expect Rayleigh-distributed envelopes[1]. However, in our framework the scatterers are not identical, they are sampled from a power law distribution of vessel diameters, and so the general trend is towards higher values of $\hat{b}$. These overall trends are shown in **Fig. 4**. As the number density and as $b$ increase within any tissue, the Burr parameters $\hat{b}$ and $\lambda$ will increase.

We note that the Burr distribution as we derive in eqn (10) has the denominator term with $(A/\lambda)^2$, and the square term can be traced back in the original derivation to the approximation of how the peak of the echo envelope increases with increasing scatterer size. This result is not precise, and can depend on the exact nature of the broadband pulse incident on a cylindrical scatterer. If we allow some perturbation of this, then the histogram formula can be written as:

$$N_{norm} = \frac{c(b-1)\left(\frac{A}{\lambda}\right)^{c-1}}{\lambda\left[\left(\frac{A}{\lambda}\right)^c + 1\right]^b}. \tag{12}$$

where $c$ is now a third parameter not constrained to 2. This is simply a more general form of the Burr distribution.[37,38] We have found that this is useful for cases with low $N_0$ that contain larger anechoic spaces. For example, if the histogram in **Fig. 6(a)** ($N_0 = 200$) is fit to eqn (12) instead of eqn (10), then we find $c = 1.7$ instead of the assumed 2, and with a higher $R^2$ (0.99 instead of 0.98). Thus, the 3-parameter Burr distribution might have general applicability to a wider range of conditions.

In the liver examples, it is plausible that fibrosis increases the number of larger scatterers (fibrotic patches), whereas fat increases greatly the number of very small (Rayleigh) scatterers. The corresponding changes in $\hat{b}$ as compared with a normal reference case are consistent with these changes in tissue. These results open the possible use of the Burr distribution parameters as biomarkers for tissue vascularity and structural composition.

*Disclosures*

The authors have no relevant financial interests in the manuscript and no other potential conflicts of interest to disclose.

*Acknowledgments*

This work was supported by National Institutes of Health grant R21EB025290. The authors also thank Terri Swanson of Pfizer Inc. for providing the RF data from their liver studies.

**Caption List**

**Fig. 1** Model of 3D convolution of a pulse with the fractal branching cylindrical fluid-filled channels in a soft tissue.

**Fig. 2** The proposed histogram function of envelope amplitudes $A$, having the form $A/(A^2 + a_{min})^b$. In (a) are normalized functions where $a_{min} = 1/2$ and the power law parameter $b$ is 3, 2.5, 2, and 1.5. In (b) are normalized functions where the power law parameter is fixed at 2.5 however $a_{min}$ is varied as 1/4, 1/2, 3/4, and 1. Vertical axis: counts (arbitrary units); horizontal axis: envelope amplitude (arbitrary units).

**Fig. 3** (a) 3D orientation of the transducer in the simulation domain. A few cylindrical branches are also shown here to clarify the random branches orientation in the domain. The length of the transducer shown here is relatively longer than that of simulation. (b) 2D view $(x-y)$ of the random distribution of cylindrical scatterers shown as black circular spots in the uniform white background, corresponding to the case of $b = 2.8$ and $N_0 = 200$.

**Fig. 4** Comparison of the random scattering fields and corresponding B-mode images for different number densities based on the power law equation for $b = 2.5$. Column (a) $N_0 = 100$, column (b) $N_0 = 200$, column (c) $N_0 = 300$, and column (d) $N_0 = 400$. A region of interest (dashed lines) is shown for analysis.

**Fig. 5** Envelope plots along the middle line for the power law parameter $b = 2.5$. The two arrows indicate the length of the ROI from which the histogram plots are obtained.

**Fig. 6** Fractal branches curve-fitting for: (a) $N_0 = 200$ and $b = 2.5$. Curve-fitting parameters: $\hat{b} = 2.341$, $\lambda = 1412$. Goodness of fit: $R^2 = 0.9869$, RMSE = 0.2088. (b) $N_0 = 300$ and $b = 2.5$. Curve-fitting parameters: $\hat{b} = 4.048$, $\lambda = 2980$. Goodness of fit: $R^2 = 0.9954$, RMSE = 0.1059. (c) $N_0 = 400$ and $b = 2.5$. Curve-fitting parameters: $\hat{b} = 5.593$, $\lambda = 4280$. Goodness of fit: $R^2 = 0.996$, RMSE = 0.08801.

**Fig. 7** Summary of simulation results using the fractal branches theory for comparison of results for fitted $\hat{b}$. Error bars for fitting parameter are also shown when $N_0 = 250$ and $b = 2.2$, 2.5, and 2.8, each resulting from 10 repetitions of the simulation.

**Fig. 8** (a) B-scan image of normal rat liver. A region of interest is selected (dashed lines) for analysis. (b) Fractal branching theory fitting to the histogram of normalized echo amplitude. Curve-fitting parameters: $\hat{b} = 2.832$, $\lambda = 407.1$. Goodness of fit: $R^2 = 0.996$, RMSE = 0.04665.

**Fig. 9** (a) B-scan image of low-fat fibrotic rat liver. A region of interest is selected (dashed lines) for analysis. (b) Fractal branching theory fitting to the histogram of normalized echo amplitude. Curve-fitting parameters: $\hat{b} = 2.184$, $\lambda = 532.5$. Goodness of fit: $R^2 = 0.9933$, RMSE = 0.03381.

**Fig 10** (a) B-scan image of high-fat fibrotic rat liver. A region of interest is selected (dashed lines) for analysis. (b) Fractal branching theory fitting to the histogram of normalized echo amplitude. Curve-fitting parameters: $\hat{b} = 3.219$, $\lambda = 871.8$. Goodness of fit: $R^2 = 0.9878$; adjusted RMSE = 0.04887.

**Table 1** Physical properties of the transducer in the simulation.

**Table 2** Averaged number of scatterers within pulse area in k-Wave simulation.

*Author Biographies and Photographs*

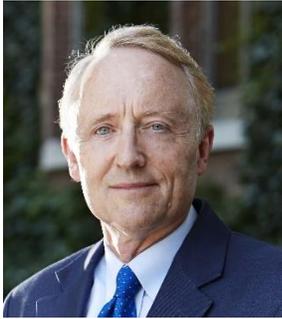

**Kevin J. Parker** is the William F. May professor of engineering at the University of Rochester. He earned his graduate degrees from Massachusetts Institute of Technology and served at the University of Rochester as department chair, director of the Rochester Center for Biomedical Ultrasound, and dean of engineering/applied sciences. He holds 26 US and 13 international patents (licensed to 25 companies), is founder of VirtualScopics, and has published 220+ journal articles. He is fellow of the IEEE, AIUM, ASA, AIMBE, and NAI.

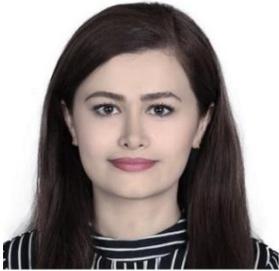

**Sedigheh S. Poul** is a PhD degree student in the department of mechanical engineering at the University of Rochester. She received her MS degree from the University of Rochester in August, 2019 and her BS degree from the University of Tehran, Iran in 2014, both in mechanical engineering. Her research interests include ultrasound elastography, tissue characterization, viscoelasticity and finite element modeling.